# MEDOC: a Python wrapper to load MEDLINE into a local MySQL database

Emeric Dynomant[1], Mathilde Gorieu[1], Hélène Perrin[1,2], Marion Denorme[1], Fabien Pichon[1], Arnaud Desfeux[1]

**Affiliation**: [1]OmicX, Seine Innopolis, 72 rue de la République, 76140, Le Petit Quevilly, France; [2]Normandie Univ, UNIROUEN, Inserm U1245 and Rouen University Hospital, Normandy Center for Genomic and Personalized Medicine, Rouen, France.

Email: ED – emeric.dynomant@omictools.com

## Abstract

Since the MEDLINE database was released, the number of documents indexed by this entity has risen every year. Several tools have been developed by the National Institutes of Health (NIH) to query this corpus of scientific publications. However, in terms of advances in big data, text-mining and data science, an option to build a local relational database containing all metadata available on MEDLINE would be truly useful to optimally exploit these resources.

MEDOC (MEdline DOwnloading Contrivance) is a Python program designed to download data on an FTP and to load all extracted information into a local MySQL database. It took MEDOC 4 days and 17 hours to load the 26 million documents available on this server onto a standard computer.

This indexed relational database allows the user to build complex and rapid queries. All fields can thus be searched for desired information, a task that is difficult to accomplish through the PubMed graphical interface. MEDOC is free and publicly available at https://github.com/MrMimic/MEDOC.

## Background

The MEDLINE database (https://www.ncbi.nlm.nih.gov/pubmed/), widely used by scientists, contains journal citations, online books, and abstracts from biomedical literature. Launched in the early 70's, it covers an extensive number of documents published since 1965 [1]. With more than 27 million citations currently indexed, it is the biggest database for scientific text mining, information extraction, natural language processing, algorithm training and reference searches, and is used daily by thousands of users.

To query this database, two different interfaces can be used, PubMed or E-utilities.

PubMed is a web interface integrating a search engine powered by the NIH [2]. Manual searches can be viewed directly by the user in their Internet browser and can be downloaded in various formats (summary, abstract, XML, PMID list, etc.). This engine, while reliable, is impractical for text mining and tasks requiring multiple searches.

To overcome this, the NIH released E-utilities [3], a group of Application Programming Interfaces (API), providing a wide range of tools for searching MEDLINE in various ways (searches by keywords, authors, etc.); *Esearch* allows basic searches in the database, *Esummary* provides summaries of documents, *Efetch* downloads full records, etc. Data can thus be retrieved automatically and parsed with informatics scripts.

However, local access to the MEDLINE database could be helpful to execute queries faster (avoiding the need to download every article to parse it) or to more easily make crossed searches between articles.

FTP access, provided by the NIH, allows users to download the full content of the database, but only in XML (eXtensible Markup Language) format. The MEDLINE database is currently divided into two repositories: the *baseline* repository (containing everything published before 2017) and the *daily update* repository (containing updates, released every 3 or 4 days, of recently published articles). These files contain all information available on MEDLINE, but are extremely large (up to 30K of articles per file) and not easily searchable without an external tool.

A relational database has the potential to be an ideal tool for both quick and complex queries. There are many models of relational databases [4], with SQL (Structured Query Language) one of the best-known and widely-used. In 2004, Oliver *et al.* converted

XML files from MEDLINE's FTP into an Oracle relational database [5]. However, this tool is now out of date for SQL technologies and file structures. It was also developed for an FTP containing less than 400 files (2 million documents), which is not only well below the current number, but is increasingly poorly-adapted given the continuously increasing publication rhythm on MEDLINE. Nonetheless, this tool was useful for downloading MEDLINE onto a server. Our goal is to allow all users to download NIH's FTP and create a local relational database from its content.

## Methods

### Database schema

The database schema presented here is that published in the article by Oliver *et al*, who constructed it from Document Type Definition (DTD) provided by the National Library of Medicine (NLM). This schema thus covers most of the fields potentially extracted from XML files downloaded onto the FTP. However, we modified the SQL script for database generation for several parameters. First, it was adapted to MySQL language instead of PostgreSQL or Oracle, this language being widely used by the bioinformatics and data science communities. Secondly, various tables were deleted (*medline_general_note*, *medline_space_flight_mission*, *medline_citation_other_abstract*) and grouped (*medline_mesh*), thus forming a 13-table relational database (Figure 1). The PubMed IDentifier (PMID) was kept as the primary or external key for all tables to link them together during requests. An index was built in the PMID field for each table, as well as in some other fields, making searching faster during querying.

### Parsing and loading software

MEDOC was constructed by coding a group of Python scripts, given that a significant proportion of the target community is able to use this programming language, thus making individual maintenance easier. Overall, this program performs eight tasks. (1) it checks if the database has already been created, and if not, builds it from the SQL sourcefile; (2) it obtains a file list from the FTP; (3) if a file has not already been added to the database, it downloads it; (4) and (5), it extracts the file from the *.gz* archive and parses it to split it into unique articles; (6) and (7), it extracts information from all articles and builds SQL queries to insert these values into the database; and (8) it removes the file and adds it to the list of files to ignore on the FTP (Figure 2).

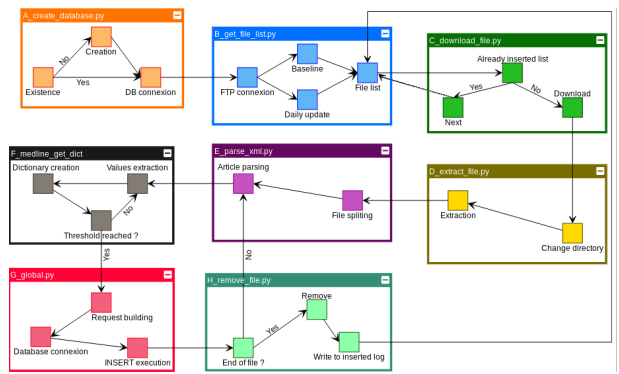

**Figure 1**: Steps executed by the program MEDOC to download MEDLINE's content and insert it into a local MySQL database.

The time consumed by parsing and loading can easily increase during this process. Thus, each action has been designed to minimize these steps (4 to 7).

The first step was to Cythonize all functions [6]. Python is a high-level programming language, easy to use and to understand, but is less efficient during execution than low-level languages such as C. The library *Cython* allows users to re-write functions originally in Python to C-like functions, thus making them faster during execution.

Second, after a file has been downloaded, it has to be parsed to extract values between XML tags. The Python library *bs4* was used for this step; more specifically its function *BeautifulSoup* [7]. The light *lxml* parser was also useful in combination with this tool, making indexation of the file faster.

Finally, SQL requests were inserted with a threshold value. This means that the program executes insertions of values into the table only when it has stored enough values to reach this threshold. This avoids connecting to the MySQL server too frequently, thus reducing the time of execution.

### Hardware configurations

All development processes and tests were performed on a standard computer, fitted with an i5-7600K 4x3.8Ghz processor and 16Go of Random Access Memory (RAM), under Ubuntu 16.04 LTS.

# Results

## Loading time and disk space use

The 1210 files to be obtained from the FTP, ranged in size from 2.5 Mo (*medline17n0670.gz*) to 46.8 Mo (*medline17n0827.gz*) for the baseline folder (896 files). It took between 0.12 and 41.21 minutes for MEDOC to download, extract and load an individual file into the local relational database depending on its size. Taken together, these files represent 22.3 Go of data. In terms of the daily update folder (279 archives), file downloading and insertion took from 0.12 min for *medline17n1153* (0.28 Mo) to 39.24 min for *medline17n1099* (60.69 Mo). Taken together, daily update files represent 4.8 Go of data.

For the database itself, 61.3 Go of disk space were used. The table with the largest number of rows was the *medline_author* (with 110,873,917 rows in total). Regarding size, the table medline_citation was the largest with 31.8 Go of disk space used.

The total loading time for all 1174 files (representing 26 million documents) was 112.85 hours (4 days and 17 hours). After loading, indexing the database for all tables took 16 minutes (index creation lines were commented during database creation to speed up insertions).

## Errors

In total, 6,050 errors were registered in the error log. In terms of the origin of the errors, 1,384 were due to data which were too long in the insert command (error code 1406), mainly in the 'initials' field. The maximum length was thus upgraded in the SQL database creation file. In addition, 4,649 errors were caused by a duplicate PMID key (error code 1062). This can be explained by a computer crash soon after initiating the insertion of values into the MySQL database. When re-started, the program attempted to add the same 4,649 values previously added. Finally, 17 errors relating to a table being full (code 1114) were due to the full disk partition, caused the above-mentioned crash.

## Database querying

The following table presents a number of SELECT commands and the time of duration of the query.

| Query | Sec | Results |
|---|---|---|
| *SELECT COUNT(pmid) FROM medline_citation ;* | *0.087* | *26,278,750* |
| *SELECT COUNT(DISTINCT(journal_title)) FROM medline_citation ;* | *0.12* | *29,848* |
| *SELECT name_of_substance FROM medline_chemical_list GROUP BY name_of_substance ;* | *0.07* | *235,182* |
| *INNER JOIN + 2 WHERE CLAUSE* | *11.84* | *265,558* |

## Conclusion

We report the development of a fast and easy-to-use tool. The relational database schema used by Oliver *et al.* has been simplified and this new Python wrapper is not only faster, but also loads almost 10-times more documents. The MySQL local database executed queries faster than E-utilities' APIs, and allows the user to build more complex requests than with PubMed's graphical interface.

Moreover, this tool has been coded to be usable on every computer, regardless of hardware configuration and with only four external dependencies to install. However, this brought several issues.

First, the quantity of RAM available on the computer could be limiting. As files on the FTP include many articles, thus becomes a major challenge to buffer and then to parse using the program to extract XML tags. Furthermore, SQL requests to insert data into the database are built with lists of fields. For tables such as *medline_citation*, which contain large fields (*text_abstract*), these lists could rapidly become difficult for the RAM buffer to handle.

For much of the time, MEDOC used between 9 and 15 Go of RAM, rising to 16 Go with 3 Go of SWAP exchange on the Linux system. Thus, for hardware configurations with less than 16 Go of RAM, the parameter *insert_command_limit* should be reduced, making the program to insert values more frequent, thereby flushing variables more often. However, it will connect to the MySQL server to execute the command more often, stretching out the time of execution.

Finally, while this program could have easily been made parallelisable to be launched on a cluster, our wish is to allow everybody to have a local version of MEDLINE, without a requirement for advanced computer skills or financial investment.

## Specifications

Project: MEDOC (MEdline DOwnloading Contrivance)

Project webpage: https://github.com/MrMimic/MEDOC

Operating system: Independent

Programming language: Python 3.5

External requirements: *Cython* (0.25.2), *pyMySQL* (0.7.11), *bs4* (4-4.6.0), *lxml* (3.8.0)

Input data: Concatenated scientific articles

Input format: XML

Output data: MySQL database cloning MEDLINE

Output format: SQL

License: None

Restrictions on use: None